\documentclass[conference]{IEEEtran}
\IEEEoverridecommandlockouts
\usepackage{cite}
\usepackage{amsmath,amssymb,amsfonts}
\usepackage{algorithmic}
\usepackage{graphicx}
\usepackage{textcomp}
\usepackage{xcolor}
\def\BibTeX{{\rm B\kern-.05em{\sc i\kern-.025em b}\kern-.08em
    T\kern-.1667em\lower.7ex\hbox{E}\kern-.125emX}}
\usepackage{url}
\usepackage{mathtools}
\usepackage{multirow}
\usepackage{makecell}
\usepackage{hologo} 
\usepackage{booktabs} 
\usepackage{tablefootnote} 
\usepackage{colortbl} 
\usepackage{xcolor}   
\usepackage{float}    
\usepackage{threeparttable} 

\DeclarePairedDelimiter\floor{\lfloor}{\rfloor}

\begin{document}

\title{Modulo-$(2^{2n}+1)$ Arithmetic via Two Parallel n-bit Residue Channels\\
\thanks{Jaberipur’s research was supported by the Brain Pool program, funded by the Ministry of Science and ICT, through the National Research Foundation of Korea (RS-2023-00263909), and Lee’s research is supported by Basic Science Research Program funded by the Ministry of Education through the National Research Foundation of Korea (NRF-2020R1I1A3063857).}
}

\author{\IEEEauthorblockN{1\textsuperscript{st} Ghassem Jaberipur}
\IEEEauthorblockA{\textit{dept. of Computer Engineering} \\
\textit{Chosun University}\\
Gwangju, Republic of Korea \\
Jaberipur@chosun.ac.kr}
\and
\IEEEauthorblockN{2\textsuperscript{nd} Bardia Nadimi}
\IEEEauthorblockA{\textit{dept. of Computer Science and Engineering} \\
\textit{University of South Florida}\\
Tampa, Florida, United States \\
bnadimi@usf.edu }
\and
\IEEEauthorblockN{3\textsuperscript{rd} Jeong-A Lee}
\IEEEauthorblockA{\textit{dept. of Computer Engineering} \\
\textit{Chosun University}\\
Gwangju, Republic of Korea \\
jalee@chosun.ac.kr}
}

\maketitle

\begin{abstract}
Augmenting the balanced residue number system moduli-set $\{m_1=2^n,m_2=2^n-1,m_3=2^n+1\}$, with the co-prime modulo $m_4=2^{2n}+1$, increases the dynamic range (DR) by around $70\%$. The Mersenne form of product $m_2  m_3  m_4=2^{4n}-1$, in the moduli-set $\{m_1,m_2,m_3,m_4\}$, leads to a very efficient reverse converter, based on the New Chinese remainder theorem. However, the double bit-width of the $m_4$ residue channel is counter-productive and jeopardizes the speed balance in $\{m_1,m_2,m_3\}$. Therefore, we decompose $m_4$ to two complex-number $n-$bit moduli $2^n\pm\sqrt{-1}$, which preserves the DR and the co-primality across the augmented moduli set. The required forward modulo-$(2^{2n}+1)$ to moduli-$(2^n\pm\sqrt{-1})$ conversion, and the reverse are immediate and cost-free. The proposed unified moduli-$(2^n\pm\sqrt{-1})$ adder and multiplier, are tested and synthesized using Spartan 7S100 FPGA. The $6-$bit look-up tables (LUT), therein, promote the LUT realizations of adders and multipliers, for $n=5$, where the DR equals $2^{25}-2^5$. However, the undertaken experiments show that to cover all the {32-}bit numbers, the power-of-two channel $m_1$ can be as wide as 12 bits with no harm to the speed balance across the five moduli.  The results also show that the moduli-$(2^5\pm\sqrt{-1})$ add and multiply operations are advantageous vs. moduli-$(2^5\pm1)$ in speed, cost, and energy measures and collectively better than those of modulo-$(2^{10}+1)$.
\end{abstract}


\section{Introduction}

Residue number systems (RNS) are the basis for the arithmetic units in several digital systems, such as cryptosystems \cite{R1}, digital signal processing \cite{R2}, image processing \cite{R3}, and recently varieties of neural network hardware accelerators \cite{R4, R5, R6, R7, R8, R9, R10, R11, R12, R13}. Fast low power/cost modular addition and multiplication, via speed-balanced residue channels with small bit-width, are the desirable and essential properties for the working RNS moduli set. As well, are the reasonably efficient forward residue generation and reverse multi-residue to binary conversion. 

The classic moduli set $T=\{2^n,2^n-1,2^n+1\}$ is quite popular and has been of utmost utility for decades. The range of uniquely representable numbers, often called the dynamic range (DR), in $T$ equals $[0,2^{3n}-2^n)$. It covers up to $18-$bit integers in $[0,262080)$, for small $n\leq6$. Such DR satisfies the working range in many applications, especially convolutional and deep neural networks (CNN and DNN). 

The corresponding modular adders and multipliers, operating on at most $6-$bit operands, are quite fast on different technological platforms. In particular, they are suitable for realization on field programmable gate arrays (FPGA) with $6-$bit look-up tables (LUT) \cite{R14}.

Another important property of $T$ is the realization of efficient residue generators and reverse conversion circuitry. Implementation of the latter, via the New Chinese remainder theorem (NCRT) \cite{R15} consists of a $2n-$bit carry-save adder (CSA) and a modulo-$(2^{2n}-1)$ addition unit. 

There are other CNN and DNN hardware accelerators that use RNS with other possibly higher moduli than those of the aforementioned $6-$bit $T$. For example, 
\begin{itemize}
    \item [1)] The three works in \cite{R5, R9, R16} use the moduli set $\{31,32,63\}$, to compute the results of their activation function, maximum pooling unit, and the processing element units. 
    \item [2)] The moduli set $\{8,63,127\}$, is used in \cite{R11} for computing the results of the convolutions or matrix products followed by the bias adjustment.
    \item [3)] The work of \cite{R13}, is based on $\{31,32,32,29,35\}$ and $\{511,512,513\}$, which offers an RNS-based convolution layer for the inferring stage.
    \item [4)] A new CNN architecture is proposed in \cite{R17}, in which convolution layers have a hardware implementation on FPGA using the RNS arithmetic with special moduli set $\{31,128,511\}$.
\end{itemize}

The moduli set, in all the above examples 1) to 4) are imbalanced, where the maximum bit-widths are $6,7,10,9$, respectively. However, the required DR can be covered via the moduli set $F=\{2^n,2^n-1,2^n+1,2^{2n}+1\}$, with smaller bit-widths $n\in\{4,4,6,5\}$, respectively. This $4-$moduli set $F$ shares the property of simple reverse conversion of $T$, since the final operation in the corresponding NCRT reverse converter is a modulo-$(2^{4n}-1)$ addition. However, the modulo-$(2^{2n}+1)$ Residue channel for $F$ is not speed-balanced with the other three moduli. Therefore, in this work, we present the new balanced adaptive moduli set $F_c=\{2^{n+p},2^n-1,2^n+1,2^n-j,2^n+j\}$, where $j=\sqrt{-1}$, and the DR is exactly equal to that of $F$, since $(2^n-j)(2^n+j)=2^{2n}+1$. Moreover, we design the required moduli-$(2^n\pm j)$ addition and multiplication schemes for LUT realization on Spartan 7S100 FPGA platform. Since the real [imaginary] parts of moduli-$(2^n\pm j)$ residues are represented in $(n+1)-$bit stored borrow [carry] format, our experiments are centered on $n=5$, for best use of the $6-$bit LUTs of the working FPGA. 

The choice of adaptive moduli set is due to the fact that the speed of Spartan 7S100 modulo-$2^{n+p}$ adders and multipliers are in balance with other moduli of $F_c$, for $p\leq n$, per the results of our experiments. Our contributions in this work are:
\begin{itemize}
    \item $n-$bit realization of modulo-$(2^{2n}+1)$ RNS arithmetic, via complex-number moduli.
    \item Efficient realizations, for $n=5$, on FPGA with $6-$bit LUTs.
    \item Taking advantage of the high performance of power-of-two modular adders and multipliers on the utilized FPGA, for setting up an adaptive moduli set.
\end{itemize}

The remaining sections of this paper discuss the following subjects. A foundational background on general RNS in Section II, followed by an in-depth examination of residue generation, adders, and multipliers for modulo-$(2^n\pm j)$ in Section III, which concludes with the critical discussion on the reverse conversion of modulo-$(2^n\pm j)$ residues to modulo-$(2^{2n}+1)$. In Section IV, we present a comparative assessment of the newly introduced complex moduli in contrast to modulo-$(2^{2n}+1)$, along with a comparison of the proposed moduli-set to other moduli sets employed in earlier DNN hardware accelerator studies. The paper concludes in Section V with a summary of findings and potential directions for future research.
\section{A Background on General RNS}	

A typical residue number system is characterized by $k$ co-prime moduli $R=\{m_1,\cdots,m_k\}$, where the DR is usually denoted by $M=m_1\times\cdots\times m_k$. A binary number $X\in[0,M)$ is represented in $R$, as $(x_1\cdots,x_k)$, where $x_i=|X|_{m_i}$ (i.e., the remainder of integer division $X/m_i$), for $1\leq i\leq k$. Execution of an arithmetic operation $Z=X\otimes Y$, where $\otimes\in\{+,\times,-\}$, and $X,Y,Z\in[0,M)$, is distributed within $k$ parallel computational residue channels (one per each modulo in $R$), such that $Z=(z_1\cdots ,z_k)$, and $z_i=|x_i\otimes y_i|_{m_i}$, for $1\leq i\leq k$.

Other arithmetic operations such as division, comparison, and sign detection cannot be performed in parallel and generally require reverse conversion of the operands to binary, while there exist some shortcut methods (e.g., \cite{R18}).

There are three major reverse conversion schemes, based on the Chinese remainder theorem (CRT); namely, the plain CRT, New CRT (NCRT), and mixed radix. The NCRT, as in (1) is of particular interest in this work, since it simplifies the reverse conversion, where $M=2^n(2^{p\times n}-1)$, which is the case in the proposed $F$ moduli set, for $p=4$.

\begin{equation} 
\label{eqn1}
\begin{split}
    X & =x_1+m_1 |\sum_{i=1}^{k-1}\mu_i(x_{i+1}-x_i)|_{M_1}, \\
    \mu_i & =(\prod_{j=2}^i m_j) |(\prod_{j=1}^i m_j )^{-1}|_{\prod_{j=i+1}^k m_j}
\end{split}
\end{equation}

\section{Modulo-$(2^n\pm j)$ Arithmetic}

Recalling the $4$-moduli set $F$, that covers $5n$-bit numbers $Z\in[0,2^{5n}-2^n)$, in this section, we provide for the forward convertor yielding $X=|Z|_{2^{2n}+1}\in[0,2^{2n}]$, followed by modulo-$(2^n\pm j)$ residue $|X|_{2^n\pm j}$ generator, adder, multiplier, and reverse double residues $|X|_{2^n\pm j}$-to-$X$ convertor. 

\subsection{Modulo-$(2^n\pm j)$ residue generator}

Let $Z=2^{4n} Z_2+2^{2n} Z_1+Z_0$, represents the $5n$-bit numbers within the dynamic range of $F$, where $Z_2=z_{5n-1}\cdots z_{4n}\in [0,2^n-1]$, $Z_1=z_{4n-1}\cdots z_{2n}\in [0,2^{2n}-1]$, and $Z_0=z_{2n-1}\cdots z_0\in [0,2^{2n}-1]$. It can be shown (See the Appendix) that $X=|Z|_{2^{2n}+1}\in [0,2^{2n}]$ is representable as in (\ref{eqn2}), where $x_{2n}=1\xrightarrow{}x_{2n-1}\cdots x_0=0$.

\begin{equation} 
\label{eqn2}
    X=x_{2n-1}\cdots x_0+\overline{x_{2n}}
\end{equation}

Let $X=2^n X_I+X_R+\overline{x_{2n}}$, where $X_R,X_I\in [0,2^n)$. Therefore, the desired complex residues of $X$ are obtained, as in (\ref{eqn3}), where $|2^n |_{2^n\pm j}=|2^n\pm j\mp j|_{2^n\pm j}=\mp j$. Note that this $X$-to-$|X|_{2^n\pm j}$ conversion, is immediate and cost-free.

\begin{equation} 
\label{eqn3}
    |X|_{2^n\pm j}=X_R+\overline{x_{2n}}\mp jX_I
\end{equation}

\subsection{Modulo-$(2^n\pm j)$ adders}

We design the required $|X+Y|_{2^n\pm j}$ adders with the practical assumption that one operand is always coming from the forward convertor as $|X|_{2^n\pm j}=X_R+\overline{x_{2n}}\mp jX_I$, per (\ref{eqn3}). The other operand is the accumulated modular sum, as the output of the modulo-$(2^n\pm j)$ adders. However, the generated carry bits $c_{R_n}$ and $c_{I_n}$ of the real and imaginary sums are transferred to the other part, since $|2^n c_{R_n}|_{2^n\pm j}=\mp jc_{R_n}$, and $|\mp j2^n c_{I_n}|_{2^n\pm j}=|\mp j\times\mp jc_{I_n}|_{2^n\pm j}=-c_{I_n}$. Therefore, we assume stored borrow (carry) representation for the real (imaginary) part of the accumulated sum, as $|Y|_{2^n\pm j}=Y_R-b_y\mp j(Y_I+c_y)$, where $Y_R,Y_I\in [0,2^n)$, and $b_y,c_y\in \{0,1\}$. Consequently, $|S|_{2^n\pm j}=|X+Y|_{2^n\pm j}$ can be unfolded as in (\ref{eqn4}), where the following propositions are in order, with due justification to follow, for the second one.

\begin{itemize}
    \item [1)] $\overline{x_{2n}}-b_y=\overline{b_y} \overline{x_{2n}}-b_y x_{2n}$,
    \item [2)] $S_R, S_I\in [0,2^n)$, $c_s=c_n+c_y x_{2n}=c_n\lor c_y x_{2n}$, and $b_s=c_{n}^{'}+b_y x_{2n}=c_{n}^{'}\lor b_y x_{2n}$, since $c_n x_{2n}=c_{n}^{'} x_{2n}=0$.
\end{itemize}

The reason, for the latter, is that $x_{2n}=1\xrightarrow{} X_R=X_I=0\xrightarrow{} 2^n c_n+S_R=X_R+Y_R+\overline{b_y} \overline{x_{2n}}=Y_R\xrightarrow{} c_n=0\xrightarrow{}c_s=c_y$, and $2^n c_{n}^{'}+S_I=X_I+Y_I+c_y \overline{x_{2n}}=Y_I\xrightarrow{} c_{n}^{'}=0\xrightarrow{}b_s=b_y$. 

\begin{itemize}
    \item Derivation of $|S|_{2^n\pm j}$:
\end{itemize}

\begin{align*}
\begin{split}
    & |S|_{2^n\pm j}= \\
    & |X_R+\overline{x_{2n}}\mp jX_I+Y_R-b_y\mp j(Y_I+c_y )|_{2^n\pm j}= \\
    & |X_R+\overline{x_{2n}}+Y_R-b_y\mp j(X_I+Y_I+c_y)|_{2^n\pm j}= \\
    & |(X_R+Y_R+\overline{b_y} \overline{x_{2n}})-b_y x_{2n}\mp \\
    & j(X_I+Y_I+c_y \overline{x_{2n}})\mp jc_y x_{2n}|_{2^n\pm j}= \\
    & |(2^n c_n+S_R)-b_y x_{2n}\mp j(2^n c_{n}^{'}+S_I)\mp jc_y x_{2n}|_{2^n\pm j}= \\
    & |S_R-c_{n}^{'}-b_y x_{2n}\mp j(S_I+c_n+c_y x_{2n})|_{2^n\pm j}\xrightarrow{}
\end{split}
\end{align*}

\begin{equation} 
\label{eqn4}
\begin{split}
    & |S|_{2^n\pm j}=S_R-b_S\mp j(S_I+c_S), \\
    & b_S=c_{n}^{'} \lor b_y x_{2n}, c_S=c_n \lor c_y x_{2n}
\end{split}
\end{equation}

\subsection{LUT realization of the modulo-$(2^n\mp j)$ adders}

Recalling $|X|_{2^n\pm j}=X_R+\overline{x_{2n}}\mp jX_I$, and $|Y|_{2^n\pm j}=Y_R-b_y\mp j(Y_I+c_y)$, let $X_I=x_{2n-1}\cdots x_n$, $X_R=x_{n-1}\cdots x_0$, $Y_I=y_{2n-1}\cdots y_n$, and $Y_R=y_{n-1}\cdots y_0$. Therefore, the required $(2^{n+1}+2)\times(2^n+1)$-bit LUTs are described as follows.

\begin{align*}
\begin{split}
    & 2^n c_n+S_R=LUT(X_R,Y_R,x_{2n},b_y)=X_R+Y_R+\overline{b_y} \overline{x_{2n}},\\
    & 2^n c_{n}^{'}+S_I=LUT(X_I,Y_I,x_{2n},c_y)=X_I+Y_I+c_y \overline{x_{2n}}, \\
    & b_s=\floor{\frac{X_I+Y_I+c_y\overline{x_{2n}}}{2^n}} \lor b_y x_{2n}, \\
    & c_s=\floor{\frac{X_R+Y_R+\overline{b_y} \overline{x_{2n}}}{2^n}} \lor c_y x_{2n}
\end{split}
\end{align*}

\begin{figure}[h]
    \centering
    \includegraphics[width=1.0\columnwidth]{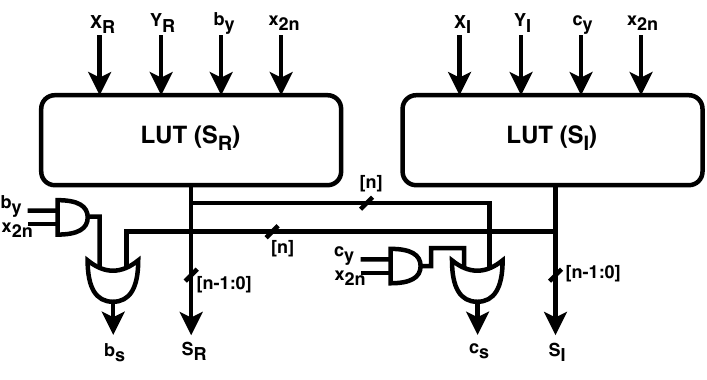}
    \caption{Proposed complex-number modulo Adder diagram.}
    \label{fig:proposed_adder}
\end{figure}

\subsection{Modulo-($2^n\pm j$) multipliers}
\label{sectionModuloMultipliers}
Let $X=x_{2n-1}\cdots x_0 + \overline{x_{2n}}$ and $Y=y_{2n-1}\cdots y_0 + \overline{y_{2n}}$ denote the multiplicand and multiplier, both as modulo-$(2^{2n}+1)$ residues generated by $F$-to-$(2^{2n}+1)$ convertor (see the Appendix). 
Therefore, recalling that $x_{2n}=1\rightarrow x_i=0$, and $y_{2n}=1\rightarrow y_i=0$, for $0\leq i<2n$, the corresponding modulo-$(2^n \pm j)$ residues are $|X|_{2^n\pm j}=\overline{x_2n}+X_R \mp jX_I=\overline{x_2n}(1+X_R\mp jX_I)$, and $|Y|_{2^n\pm j}=\overline{y_{2n}}+Y_R\mp jY_I=\overline{y_{2n}}(1+Y_R\mp jY_I)$. Consequently, $|X\times Y|_{2^n\pm j}$ is expressed as follows.

\begin{equation} 
\label{eqn5}
\begin{split}
    & |X\times Y|_{2^n\pm j}= \\
    & |\overline{x_{2n}}  \overline{y_{2n}}(1+X_R\mp jX_I)(1+Y_R\mp jY_I)|_{2^n\pm j}= \\
    & \overline{x_{2n}}  \overline{y_{2n}} |(1+X_R)(1+Y_R)\mp j(1+X_R) Y_I\mp \\
    & jX_I (1+Y_R) - X_I\times Y_I |_{2^n\mp j}
\end{split}
\end{equation}

Let the four product expressions, within \ref{eqn5}, be expressed as 1 to 4, below:

\begin{itemize}
    \item [1)] $(1+X_R)(1+Y_R)=2^{2n}c+2^n H_{RR}+L_{RR}\rightarrow$\\$|(1+X_R)(1+Y_R)|_{2^n\pm j}=|L_{RR}-c\mp jH_{RR} |_{2^n\mp j}$
    \item [2)] $(1+X_R) Y_I=2^n H_{RI}+L_{RI}\rightarrow |(1+X_R) Y_I|_{2^n\pm j}=L_{RI}\mp jH_{RI}$
    \item [3)] $X_I (1+Y_R)=2^n H_{IR}+L_{IR}\rightarrow |X_I (1+Y_R)|_{2^n\pm j}=L_{IR}\mp jH_{IR}$
    \item [4)] $X_I\times Y_I=2^n H_{II}+L_{II}\rightarrow 2^{2n}-1-X_I\times Y_I=2^{2n}-1-(2^n H_{II}+L_{II})=2^n \overline{H_{II}}+\overline{L_{II}}\times |-X_I\times Y_I|_{2^n\pm j}=|2^{2n}+1-X_I\times Y_I|_{2^n\pm j}=|2+2^n \overline{H_{II}}+\overline{L_{II}}|_{2^n\pm j}\rightarrow |-X_I\times Y_I|_{2^n\pm j}=|2+\overline{L_{II}}\mp j\overline{H_{II}}|_{2^n\pm j}$
\end{itemize}

Therefore,

\begin{align*}
\begin{split}
    & |X\times Y|_{2^n\pm j}= \\
    & \overline{x_{2n}} \overline{y_{2n}} |L_{RR}-c\mp jH_{RR}\mp j(L_{RI}\mp jH_{RI})\mp j(L_{IR}\mp jH_{IR}) \\
    & +2+\overline{L_{II}}\mp j\overline{H_{II}}|_{2^n\pm j}=\overline{x_{2n}}  \overline{y_{2n}} |L_{RR}-c+\overline{L_{II}}-H_{RI}- \\
    & H_{IR}+2\mp j(H_{RR}+L_{RI}+L_{IR}+\overline{H_{II}})|_{2^n\pm j}
\end{split}
\end{align*}

Given that $|-H_{RI}-H_{IR}|_{2^n\pm j}=|2(2^n\pm j)-H_{RI}-H_{IR}|_{2^n\pm j}=|\overline{H_{RI}}+\overline{H_{IR}}+2\pm 2j|_{2^n\pm j}$, further elaboration leads to:

$|X\times Y|_{2^n\pm j}=|L_{RR}+\overline{L_{II}}+\overline{H_{RI}}+\overline{H_{IR}}+1-c+3\pm 2j\mp j(H_{RR}+L_{RI}+L_{IR}+\overline{H_{II}})|_{2^n\pm j}=|L_{RR}+\overline{L_{II}}+\overline{H_{RI}}+\overline{H_{IR}} +\overline{c}+3\mp j(H_{RR}+L_{RI}+L_{IR}+\overline{H_{II}}-2)|_{2^n\pm j}\rightarrow |X\times Y|_{2^n\pm j}=|R\mp jI|_{2^n\pm j}$, where

\begin{equation} 
\label{eqn6}
\begin{split}
    & R=L_{RR}+\overline{L_{II}}+\overline{H_{RI}}+\overline{H_{IR}}+\overline{c}+3, \\
    & I=H_{RR}+L_{RI}+L_{IR}+\overline{H_{II}}-2
\end{split}
\end{equation}

\subsection{LUT realization of $|X\times Y|_{2^n\mp j}$}
The required LUTs for the four $(H, L)$ pairs in 1) to 4), of Section \ref{sectionModuloMultipliers}, are described as $LUT_1$ to $LUT_4$, below.

\begin{align*}
\begin{split}
    & LUT_1 ((1+X_R)(1+Y_R))=2^{2n}c+2^n H_{RR}+L_{RR}.\\
    & LUT_2 ((1+X_R) Y_I )=2^n H_{RI}+L_{RI}.\\
    & LUT_3 (X_I (1+Y_R))=2^n H_{IR}+L_{IR}.\\
    & LUT_4 (X_I\times Y_I )=2^n H_{II}+L_{II}. 
\end{split}
\end{align*}

Trusting $R-3=L_{RR}+\overline{L_{II}}+\overline{H_{RI}}+\overline{H_{IR}}+\overline{c}$ to a $n-$bit $(4; 2)$ compressor, leads to $|R-3|_{2^n\mp j}=|2^n v_n+ 2^n c_n+ \hat{V} + \hat{U} |_{2^n\mp j}=|\hat{V}+\hat{U} \pm j(v_n+c_n)|_{2^n\mp j}$  , where $\hat{V} = v_{n-1}\cdots v_1 0$, and  $\hat{U} = u_{n-1}\cdots u_0$.
Similar compression of $I+2+c_n+v_n=H_{RR}+L_{RI}+L_{IR}+\overline{H_{II}}+c_n+v_n$, results in $|I+2+c_n+v_n |_{2^n\mp j}=|2^n v^{'}_n +2^n c^{'}_n+\hat{V^{'}} + \hat{U^{'}} |_{2^n\mp j}=|\hat{V^{'}} + \hat{U^{'}} \pm j(v_n^{'}+c_n^{'} )|_{2^n\mp j}$, where $\hat{V^{'}} = v_{n-1}^{'}\cdots v_1^{'} v_0^{'}$,$\hat{U^{'}} = u_{n-1}^{'}\cdots u_0^{'}$. Therefore,

\begin{align*}
\begin{split}
|R\mp jI|_{2^n\pm j} &=
    |\hat{V}+\hat{U} \mp j(v_n+c_n)+3\mp \\
    & j(\hat{V^{'}} + \hat{U^{'}} \mp j(v_n^{'}+c_n^{'})-2-c_n-v_n )|_{2^n\pm j}= \\ 
    & |\hat{V}+\hat{U} - v_n^{'}-c_n^{'}+3\mp \\
    & j(\hat{V^{'}} + \hat{U^{'}} + 2^n-2-2^n )|_{2^n\pm j}= \\
    & |\hat{V}+\hat{U} + 1-v_n^{'}+1-c_n^{'}+1\mp \\
    & j(\hat{V^{'}} + \hat{U^{'}} + 2^n-2\pm j)|_{2^n\pm j}= \\
\end{split}
\end{align*}
\begin{flushleft}
$|\hat{V}+\hat{U} +\overline{v_n^{'}}+\overline{c_n^{'}}+2\mp j(\hat{V^{'}} + \hat{U^{'}} + 2^n-2)|_{2^n\pm j}\rightarrow$
\end{flushleft}
\vspace*{-4pt}
\begin{equation} 
\label{eqn7}
\begin{split}
    & |R\mp jI|_{2^n\pm j}=\\
    & |\hat{V}+\hat{U} +\overline{v_n^{'}}+\overline{c_n^{'}}+2\mp j(\hat{V^{'}} + \hat{U^{'}} + 2^n-2)|_{2^n\pm j}
\end{split}
\end{equation}

Let $L_{RR}=d_{n-1}\cdots d_0$, $H_{RR}=e_{n-1}\cdots e_0$, $\overline{L_{II}}=f_{n-1}\cdots f_0$, $\overline{H_{II}}=g_{n-1}\cdots g_0$,$L_{RI}=h_{n-1}\cdots h_0$, $\overline{H_{RI}}=l_{n-1}\cdots l_0$, $L_{IR}=r_{n-1}\cdots r_0$, $\overline{H_{IR}}=t_{n-1}\cdots t_0$.\\
The LUT realization of the $n-$bit adders that yield the product $P_R-b_P\mp j(P_I+c_P)$ are as follows.\\

\noindent $P_R=LUT(W,Z)=|W+Z+1|_{2^n}, c_P=\frac{W+Z+1}{2^n}$\\
$P_I=LUT(W^{'},Z^{'})=|W^{'}+Z^{'} |_{2^n}, b_P=\frac{W^{'}+Z^{'}}{2^n}.$

\begin{table}[t]
\centering
\caption{$(4; 2)$ COMPRESSORS FOR THE MULTIPLIER}
\resizebox{\columnwidth}{!}{
\begin{tabular}{|c|c||c|c|c|c||c|c|c|c|c|}
\hline
 & $L_{RR}$ & $d_{n-1}$ & $\cdots$ & $d_1$ & $d_0$ & $H_{RR}$ & $e_{n-1}$ & $\cdots$ & $e_1$ & $e_0$ \\ \cline{2-11}
 $(4;2)$ & $\overline{L_{II}}$ & $f_{n-1}$ & $\cdots$ & $f_1$ & $f_0$ & $L_{RI}$ & $h_{n-1}$ & $\cdots$ & $c_1$ & $c_0$ \\ \cline{2-11}
compressors & $\overline{H_{RI}}$ & $l_{n-1}$ & $\cdots$ & $l_1$ & $l_0$ & $L_{IR}$ & $r_{n-1}$ & $\cdots$ & $r_1$ & $r_0$ \\ \cline{2-11}
$6\Delta G$ & $\overline{H_{IR}}$ & $t_{n-1}$ & $\cdots$ & $t_1$ & $t_0$ & $\overline{H_{II}}$ & $g_{n-1}$ & $\cdots$ & $g_1$ & $g_0$ \\ \cline{2-11}
\multirow{3}{*}{} & & & & & $\overline{c}$ & & & & & $c_n$ \\ \cline{2-11}
    & & & & & & & & & & $v_n$ \\ \Xhline{5\arrayrulewidth}
Carry-save & $\hat{U}$ & $u_{n-1}$ & $\cdots$ & $u_1$ & $u_0$ & $\hat{U}^{'}$ & $u_{n-1}^{'}$ & $\cdots$ & $u_1^{'}$ & $u_0^{'}$ \\ \cline{2-11}
adders & $\hat{V} + \overline{v_n^{'}}$ & $v_{n-1}$ & $\cdots$ & $v_1$ & $\overline{v_n^{'}}$ & $\hat{V^{'}}$ & $v_{n-1}^{'}$ & $\cdots$ & $v_1^{'}$ & $v_0^{'}$ \\ \cline{2-11}
$2\Delta G$& & & & & $\overline{c_n^{'}}$ & $2^n-2$ & $1$ & $\cdots$ & $1$ & \\ \Xhline{5\arrayrulewidth}
$n-$bit & $W$ & $w_{n-1}$ & $\cdots$ & $w_1$ & $w_0$ & $W^{'}$ & $w_{n-1}^{'}$ & $\cdots$ & $w_1^{'}$ & $w_0^{'}$ \\ \cline{2-11}
adders & $Z$ & $z_{n-1}$ & $\cdots$ & $z_1$ & $\overline{w_n^{'}}$ & $Z^{'}$ & $z_{n-1}^{'}$ & $\cdots$ & $z_1^{'}$ & $z_0^{'}$ \\ \hline
\multirow{2}{*}{$|R \pm j I|_{2^n \mp j}$} & $P_R$ & $p_{n-1}$ & $\cdots$ & $p_1$ & $p_0$ & $P_I$ & $p_{n-1}^{'}$ & $\cdots$ & $p_1^{'}$ & $p_0^{'}$ \\ \cline{2-11}
    & & & & & $-b_P$ & & & & & $c_P$ \\ \Xhline{5\arrayrulewidth}
 & & & & & & \multicolumn{5}{c}{$\times \pm j$} \\ \hline
\end{tabular}
}
\label{tab:compressors}
\end{table}

\begin{figure}[h]
    \centering
    \includegraphics[width=1.0\columnwidth]{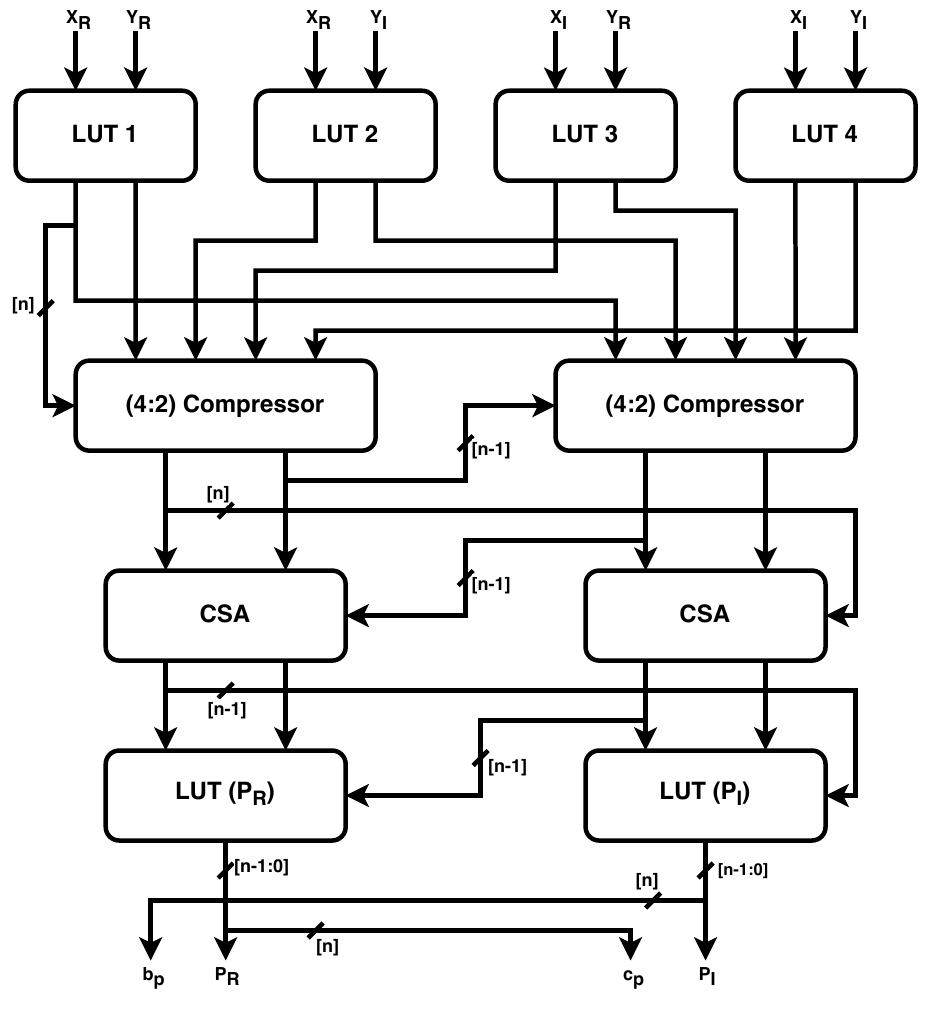}
    \caption{Proposed complex-number modulo Multiplier diagram.}
    \label{fig:2}
\end{figure}

\subsection{Immediate and cost-free Reverse moduli-$(2^n\pm j)$ to modulo-$(2^{2n}+1)$ convertor}
The reverse conversion task is meant to derive the $(2n+1)-$bit $S [P]$ from the two conjugate residues of the modulo-$(2^n\pm j)$ sum [product], such that $|S|_{2^n\pm j}=S_R-b_S\mp j(S_I+c_S)$ [$|P|_{2^n\pm j}=P_R-b_P\mp j(P_I+c_P)$]. \\
The modulo-$(2^{2n}+1)$ sum $S$ can be expressed as $S=2^n(s_I+c_s)+s_R-b_s$, which is easily verifiable. 
However, to obtain a clean-cut $(2n+1)-$bit result for $S$, we derive $S=|S|_{2^{2n}+1}$, as follows, where similar undertakings apply to $P$.\\
$S=|2^n(s_I+c_s)+s_R-b_s|_{2^{2n}+1}=$ \\
$|2^{2n}+2^n s_I+s_R+2^n c_s+\overline{b_s}|_{2^{2n}+1}$, which can be obtained via a sparse modulo-$(2^{2n}+1)$ adder, where the main operand is the $(2n+1)-$bit $2^{2n}+2^n s_I+s_R$ and the sparse one contains only two bits in positions $0(\overline{b_s})$, and $n(c_s)$. \\
However, since this result is used in a further reverse conversion, along the moduli-$(2^n\pm 1)$ and the power-of-two modulo, it can be used as $S=2^n (s_I+c_s)+s_R-b_s$, where no cost nor delay is imposed regarding the conversion of the two complex-number residues to modulo-$(2^{2n}+1)$ one. 
\vspace*{-8pt}

\section{Evaluation And Comparison}
Realization of the proposed adders and multipliers on FPGAs represents a significant step towards validating their practical applicability and performance characteristics. 
This section delves into the implementation details of these arithmetic units and their implementation on the Spartan-7S100 FPGA platform using the Vivado FPGA synthesis tool. 
Through this FPGA realization, we aim to demonstrate the practical merits of the adders and multipliers, highlighting their suitability for high-performance computing applications that demand reliable and efficient arithmetic operations. 
To ensure a fair and objective comparison of the proposed adders and multipliers with existing solutions, all similar works have been implemented on the same platform. 
This uniformity in the implementation platform allows for a direct and equitable evaluation of each design's performance, resource utilization, and power efficiency. 
By adopting the Spartan-7S100 FPGA across all designs, we eliminate variables that could arise from differences in hardware capabilities, synthesis optimizations, or architectural efficiencies inherent to different FPGA models. 
This approach guarantees that the comparative analysis focuses solely on the merits of the design methodologies and the inherent efficiencies of the proposed arithmetic units.\\
Table \ref{tab:moduli_performance} presents the delay, area, and power measures of the proposed moduli compared to the superseded modulo. 
As indicated in the table, both moduli-$(2^5\pm j)$ adder and multiplier outperform the moduli-$(2^{10}+1)$ counterparts regarding all the three figures of merit. 
However, concerning the multipliers’ area measures, note that when comparing the area of two FPGA designs, it's important to consider not just the number of utilized LUTs, but also the inclusion of other generic components like digital signal processor (DSP) blocks. 
In this case, while the second design uses fewer LUTs, the engagement of a DSP block can significantly impact the overall area. 
The reason is that the FPGA DSP blocks are universal units designed for efficient arithmetic operations and typically exhibit resource consumption overhead in comparison to specialized units. 

\begin{table}[ht]
\centering
\caption{Figures of Merit for Double Complex-Number Moduli Versus the Equivalent Double Bit-Width Modulo}
\resizebox{\columnwidth}{!}{
\begin{threeparttable}
\begin{tabular}{|c|c|c|c|c|c|c|c|}
\hline
\multirow{3}{*}{\textbf{Moduli}} & \multicolumn{4}{c|}{\textbf{Adder}} & \multicolumn{3}{c|}{\textbf{Multiplier}} \\ \cline{2-8} 
& \textbf{Delay} & \textbf{Area} & \textbf{Power} & \multirow{2}{*}{\textbf{PDP}} & \textbf{Delay} & \textbf{Area} & \textbf{Power} \\ 
& \textbf{(ns)} & \textbf{(LUT)} & \textbf{($\mu$W)} &  & \textbf{(ns)} & \textbf{(LUT)} & \textbf{($\mu$W)} \\ \hline
$2^5 \pm j$    & 9.562 & 32  & 14.162 & 135 & 20.402 & 428 & 14.586 \\ \hline
$2^{10} + 1$   & 16.19 & 67  & 10.515 & 170 & 23.72  & 214+1\tnote{*} & 15.789 \\
\hline
\end{tabular}
\begin{tablenotes}
\footnotesize
\item[*] DSP.
\end{tablenotes}
\end{threeparttable}
}
\label{tab:moduli_performance}
\end{table}

\begin{figure}[H]
    \centering
    \includegraphics[width=1.0\columnwidth]{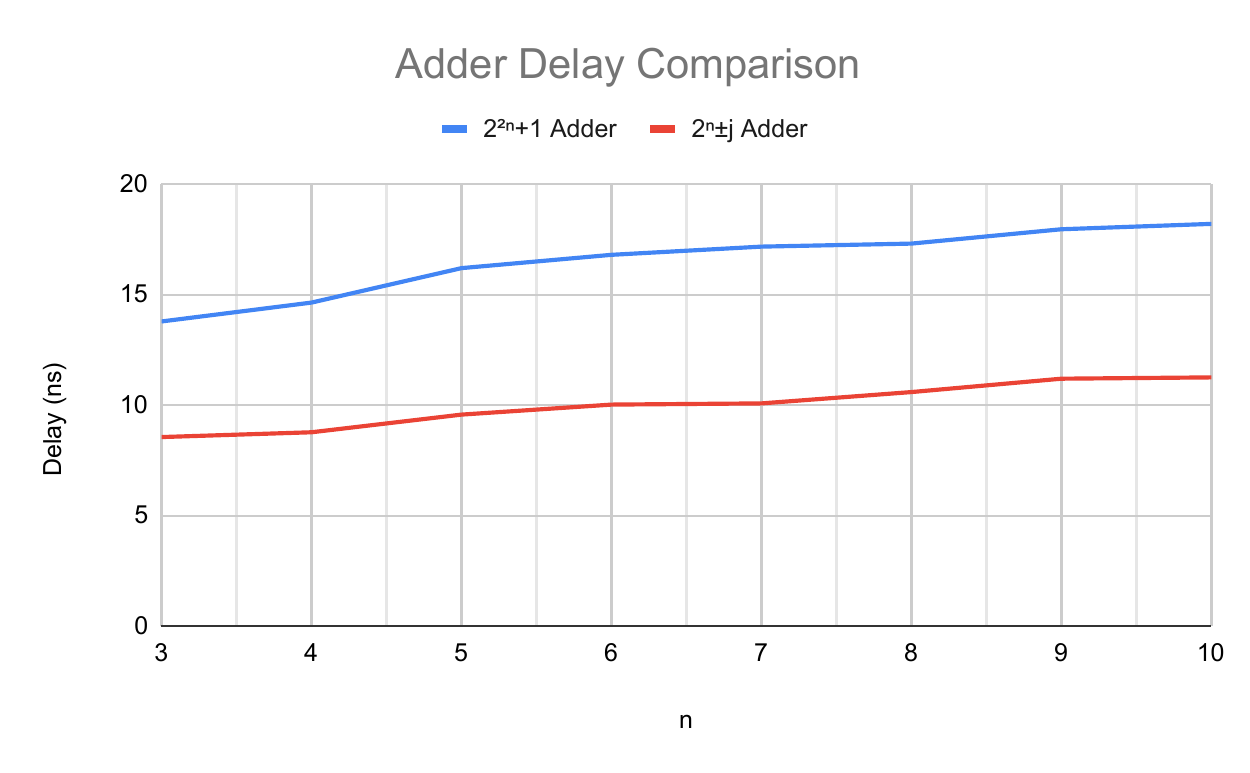}
    \caption{Complex-number Modulo vs modulo-$(2^{2n}+1)$ adder: Delay comparison.}
    \label{fig:3}
\end{figure}

\begin{figure}[H]
    \centering
    \includegraphics[width=1.0\columnwidth]{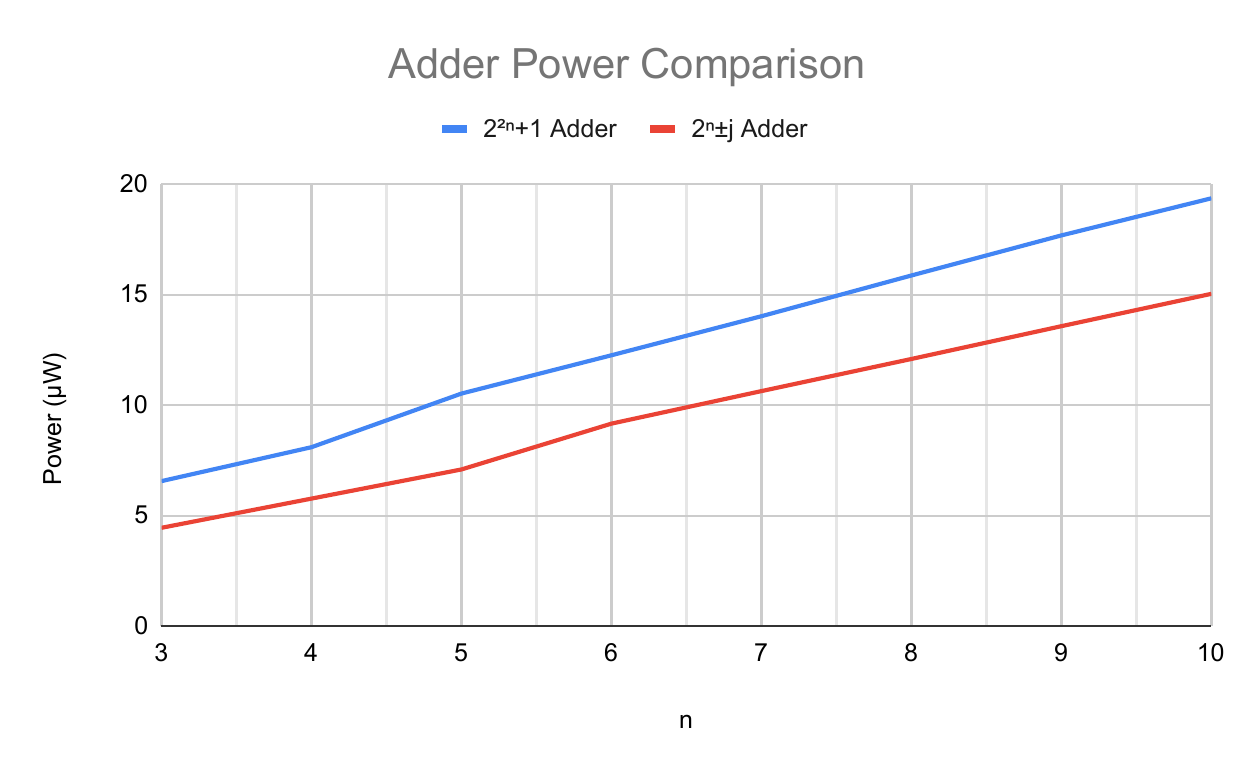}
    \caption{Complex-number Modulo vs modulo-$(2^{2n}+1)$ adder: Power comparison.}
    \label{fig:4}
\end{figure}

\begin{figure}[H]
    \centering
    \includegraphics[width=1.0\columnwidth]{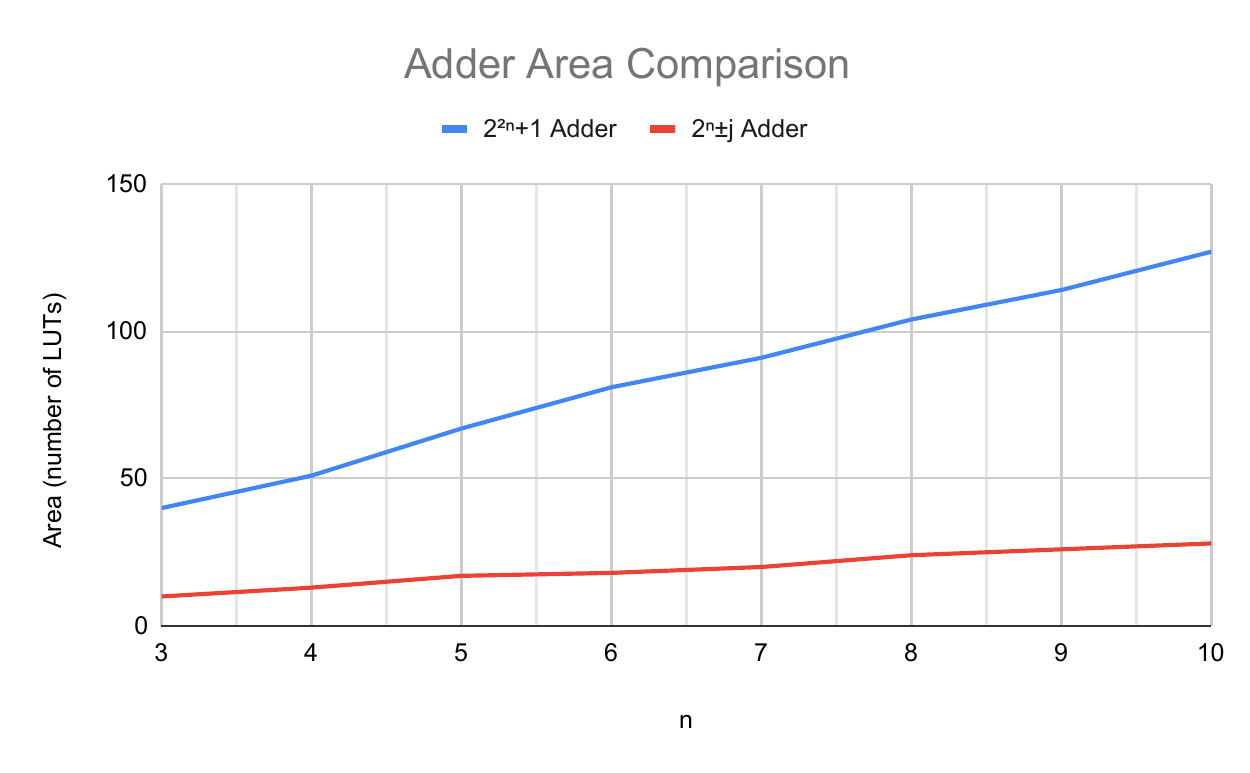}
    \caption{Complex-number Modulo vs modulo-$(2^{2n}+1)$ adder: Area comparison.}
    \label{fig:5}
\end{figure}

\begin{figure}[H]
    \centering
    \includegraphics[width=1.0\columnwidth]{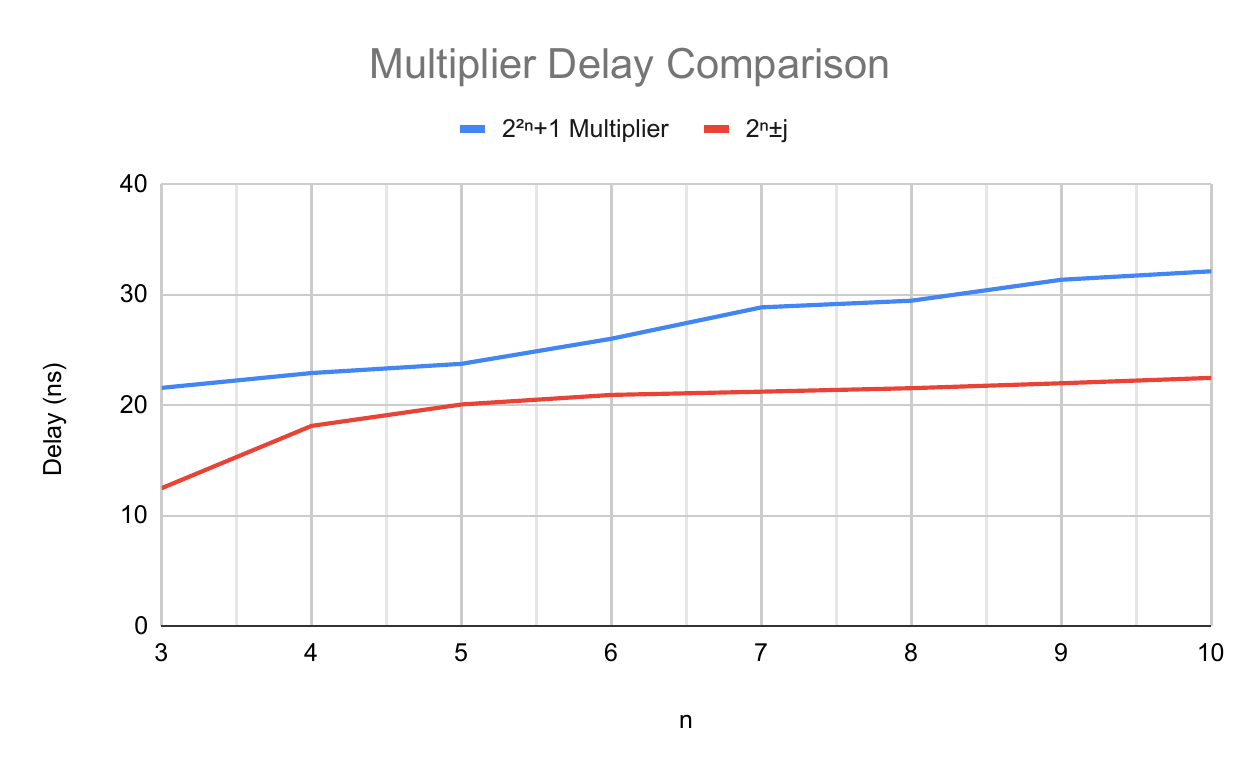}
    \caption{Complex-number Modulo vs modulo-$(2^{2n}+1)$ multiplier: Delay comparison.}
    \label{fig:6}
\end{figure}

\begin{figure}[H]
    \centering
    \includegraphics[width=1.0\columnwidth]{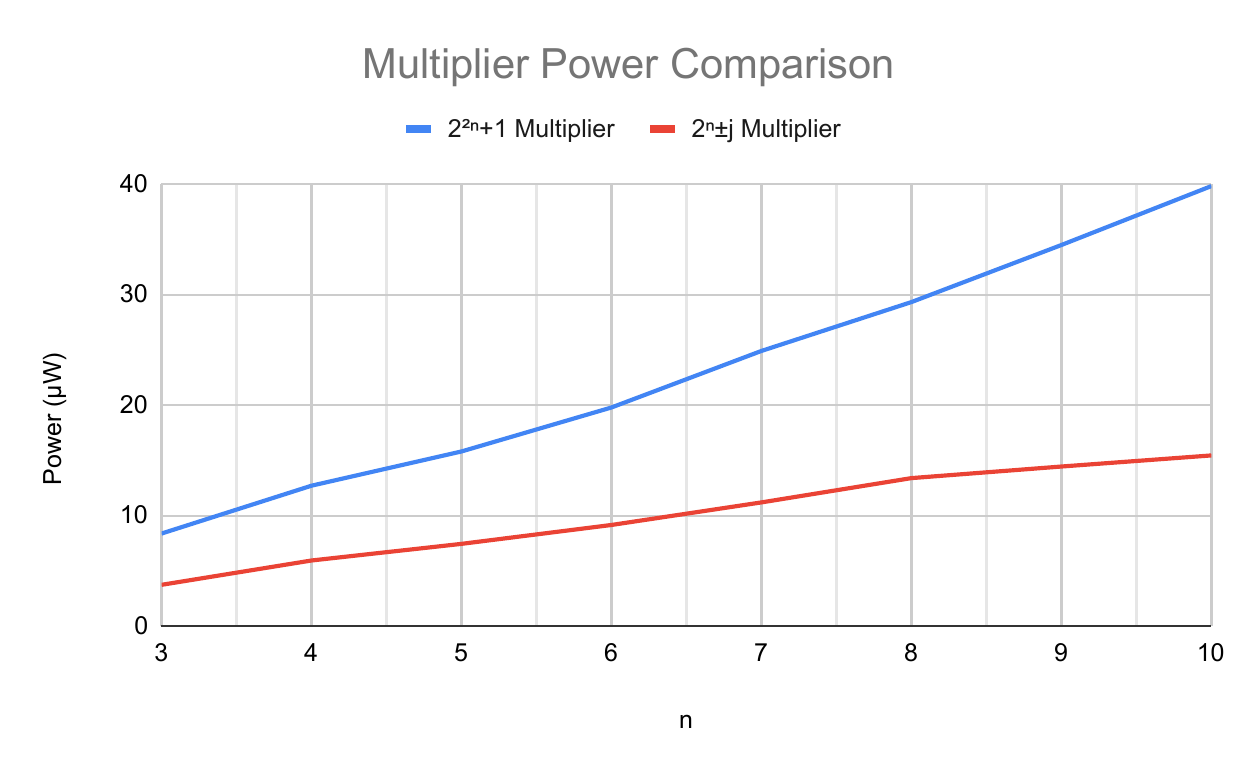}
    \caption{Complex-number Modulo vs modulo-$(2^{2n}+1)$ multiplier: Power comparison.}
    \label{fig:7}
\end{figure}

\begin{figure}[H]
    \centering
    \includegraphics[width=1.0\columnwidth]{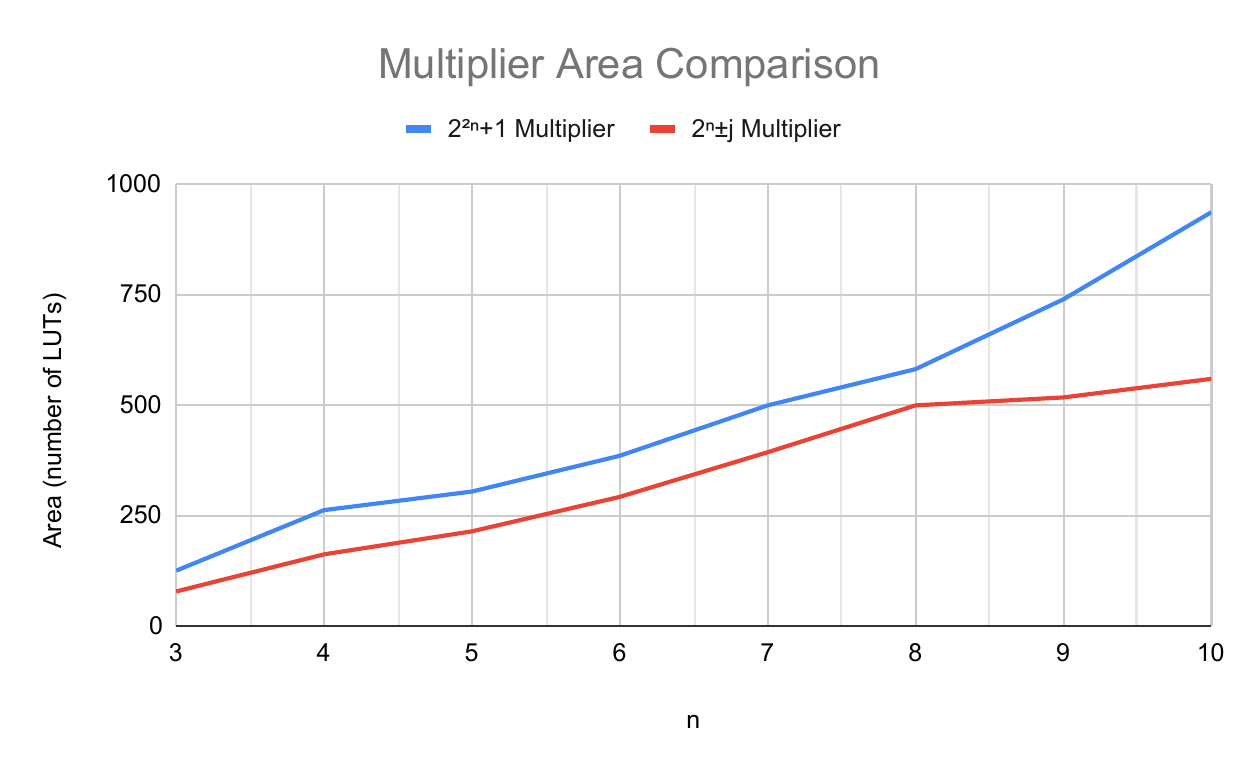}
    \caption{Complex-number Modulo vs modulo-$(2^{2n}+1)$ multiplier: Area comparison.}
    \label{fig:8}
\end{figure}

To show the substantial features of the proposed complex-number modulo adders and multipliers for a range of bit-width values, we provide for the Figures \ref{fig:3} to \ref{fig:8}. 
The six plots therein illustrate the delay, area consumption and power dissipation of the proposed designs and those of modulo-$(2^{2n}+1)$, for $3\leq n\leq 10$. 
These figures clearly demonstrate that the proposed moduli significantly outperform the counterpart in terms of delay, area, and power consumption.\\
Additionally, concerning the area measurements, for modulo-$(2^{2n}+1)$ when $n\geq5$ and for the proposed moduli-$(2^n\pm j)$ when $n\geq 9$, the area depicted in the figures has been converted to all LUTs based on the average size of Digital Signal Processors (DSPs) derived from the synthesis results.
This conversion was performed solely for the purpose of comparison.

As anticipated, the synthesis results reveal that the slowest residue channel is attributed to modulo-$(2^n+1)$. Tables \ref{tab:3} to \ref{tab:8} provide a comparison of the slowest channel delay, total area, and power consumption between the proposed replacement moduli-sets (RP1) and those utilized in previous RNS-DNN studies. 

\begin{table}[H]
\centering
\caption{Moduli-Set Replacements for RES-DNN \cite{R5}, RNSIM \cite{R9}, and RNS-PM \cite{R16}}
\resizebox{\columnwidth}{!}{
\begin{tabular}{|c|c|c|c|c|c|}
\hline
\multirow{2}{*}{\textbf{Moduli-Sets}} & \multicolumn{5}{c|}{\textbf{Adder}} \\ \cline{2-6}
& \textbf{Delay (ns)} & \textbf{Area (LUT)} & \textbf{Power ($\mu$W)} & \textbf{PDP} & \textbf{DR} \\ \hline
$\{31, 32, 63\}$ & 11.791 & 44 & 12.618 & 149 & 62,496 \\ \hline
RP1: $\{7, 9, 16, 8 \mp j\}$ & 8.546 & 38 & 15.596 & 133 & 65,520 \\ \hline
\rowcolor{lightgray} 
RP2: $\{15, 64, 8 \mp j\}$ & 8.851 & 34 & 14.312 & 127 & 62,400 \\ \hline \hline
& \multicolumn{5}{c|}{\textbf{Multiplier}} \\ \hline
$\{31, 32, 63\}$ & 21.938 & 15.684 & 171 & 344 & 62,496 \\ \hline
RP1: $\{7, 9, 16, 8 \mp j\}$ & 12.464 & 14.251 & 197 & 178 & 65,520 \\ \hline
RP2: $\{15, 64, 8 \mp j\}$ & 12.464 & 15.793 & 220 & 197 & 62,400 \\ \hline
\end{tabular}
}
\label{tab:3}
\end{table}

\begin{table}[H]
\centering
\caption{Moduli-Set Replacements for RES-DNN \cite{R4}}
\resizebox{\columnwidth}{!}{
\begin{tabular}{|c|c|c|c|c|c|}
\hline
\multirow{2}{*}{\textbf{Moduli-Sets}} & \multicolumn{5}{c|}{\textbf{Adder}} \\ \cline{2-6}
& \textbf{Delay (ns)} & \textbf{Area (LUT)} & \textbf{Power ($\mu$W)} & \textbf{PDP} & \textbf{DR} \\ \hline
$\{63, 64, 65\}$                      & 13.777              & 73                  & 15.934                  & 220          & 262,080     \\ \hline
RP1: $\{7, 9, 16, 16 \mp j\}$         & 8.763               & 44                  & 18.238                  & 160          & 259,056     \\ \hline
\rowcolor{lightgray} 
RP2: $\{15, 128, 16 \mp j\}$          & 8.851               & 45                  & 18.178                  & 160          & 493,440     \\ \hline \hline
& \multicolumn{5}{c|}{\textbf{Multiplier}} \\ \hline
$\{63, 64, 65\}$                      & 21.938              & 239                 & 19.452                  & 427          & 262,080     \\ \hline
RP1: $\{7, 9, 16, 16 \mp j\}$         & 18.102              & 365                 & 18.653                  & 338          & 259,056     \\ \hline
RP2: $\{15, 128, 16 \mp j\}$          & 18.102              & 396                 & 21.25                   & 385          & 493,440     \\ \hline
\end{tabular}
}
\label{tab:4}
\end{table}

\begin{table}[H]
\centering
\caption{Moduli-Set Replacements for RES-DNN \cite{R11}}
\resizebox{\columnwidth}{!}{
\begin{tabular}{|c|c|c|c|c|c|}
\hline
\multirow{2}{*}{\textbf{Moduli-Sets}} & \multicolumn{5}{c|}{\textbf{Adder}} \\ \cline{2-6}
& \textbf{Delay (ns)} & \textbf{Area (LUT)} & \textbf{Power ($\mu$W)} & \textbf{PDP} & \textbf{DR} \\ \hline
$\{8, 63, 127\}$                      & 11.803              & 56                  & 14.661                  & 173          & 64,008      \\ \hline
RP1: $\{7, 9, 16, 8 \mp j\}$          & 8.546               & 38                  & 15.596                  & 133          & 65,520      \\ \hline
\rowcolor{lightgray} 
RP2: $\{15, 128, 8 \mp j\}$           & 8.851               & 39                  & 15.536                  & 138          & 124,800     \\ \hline \hline
& \multicolumn{5}{c|}{\textbf{Multiplier}} \\ \hline
$\{8, 63, 127\}$                      & 22.135              & 211                 & 16.32                   & 361          & 64,008      \\ \hline
RP1: $\{7, 9, 16, 8 \mp j\}$          & 12.464              & 134                 & 13.022                  & 162          & 65,520      \\ \hline
RP2: $\{15, 128, 8 \mp j\}$           & 12.464              & 228                 & 16.848                  & 210          & 124,800     \\ \hline
\end{tabular}
}
\label{tab:5}
\end{table}

\begin{table}[H]
\centering
\caption{Moduli-Set Replacements for RES-DNN \cite{R12}}
\resizebox{\columnwidth}{!}{
\begin{threeparttable}
\begin{tabular}{|c|c|c|c|c|c|}
\hline
\multirow{2}{*}{\textbf{Moduli-Sets}} & \multicolumn{5}{c|}{\textbf{Adder}} \\ \cline{2-6}
& \textbf{Delay (ns)} & \textbf{Area (LUT)} & \textbf{Power ($\mu$W)} & \textbf{PDP} & \textbf{DR} \\ \hline
$\{3, 5, 7, 11, 13, 16, 17, 19, 23\}$ & 12.407              & 125                 & 29.315                  & 364          & 1,784,742,960 \\ \hline
RP1: $\{63, 65, 128, 64 \mp j\}$      & 13.777              & 110                 & 35.466                  & 489          & 2,147,483,520 \\ \hline
\rowcolor{lightgray} 
RP2: $\{15, 31, 1024, 64 \mp j\}$     & 10.015              & 72+1\tnote{*}       & 30.831                  & 309          & 1,950,827,520 \\ \hline 
& \multicolumn{5}{c|}{\textbf{Multiplier}} \\ \hline
$\{3, 5, 7, 11, 13, 16, 17, 19, 23\}$ & 24.845              & 348                 & 34.232                  & 850          & 1,784,742,960 \\ \hline
RP1: $\{63, 65, 128, 64 \mp j\}$      & 21.938              & 831                 & 38.781                  & 851          & 2,147,483,520 \\ \hline
RP2: $\{15, 31, 1024, 64 \mp j\}$     & 20.906              & 690+1\tnote{*}      & 36.625                  & 766          & 1,950,827,520 \\ \hline
\end{tabular}
\begin{tablenotes}
\footnotesize
\item[*] DSP.
\end{tablenotes}
\end{threeparttable}
}
\label{tab:6}
\end{table}

\begin{table}[H]
\centering
\caption{Moduli-Set Replacements for RES-DNN \cite{R13}}
\resizebox{\columnwidth}{!}{
\begin{threeparttable}
\begin{tabular}{|c|c|c|c|c|c|}
\hline
\multirow{2}{*}{\textbf{Moduli-Sets}} & \multicolumn{5}{c|}{\textbf{Adder}} \\ \cline{2-6}
& \textbf{Delay (ns)} & \textbf{Area (LUT)} & \textbf{Power ($\mu$W)} & \textbf{PDP} & \textbf{DR} \\ \hline
L: $\{32, 31, 33, 29, 35\}$           & 15.297              & 131                 & 24.156                  & 370          & 33,227,040 \\ \hline
RP1: $\{31, 32, 33, 32 \mp j\}$       & 12.65               & 88                  & 26.487                  & 335          & 33,554,400 \\ \hline
\rowcolor{lightgray} 
RP2: $\{15, 31, 128, 32 \mp j\}$      & 9.562               & 64                  & 24.503                  & 234          & 61,008,000 \\ \hline \hline
H: $\{512, 511, 513\}$                & 14.894              & 107                 & 24.074                  & 359          & 134,217,216 \\ \hline
RP1: $\{31, 32, 33, 64 \mp j\}$       & 12.65               & 92                  & 30.633                  & 388          & 134,119,392 \\ \hline
\rowcolor{lightgray} 
RP2: $\{15, 31, 128, 64 \mp j\}$      & 10.015              & 69                  & 28.649                  & 287          & 243,853,440 \\ \hline \hline
& \multicolumn{5}{c|}{\textbf{Multiplier}} \\ \hline
L: $\{32, 31, 33, 29, 35\}$           & 21.09               & 397                 & 29.932                  & 631          & 33,227,040 \\ \hline
RP1: $\{31, 32, 33, 32 \mp j\}$       & 21.09               & 619                 & 30.701                  & 647          & 33,554,400 \\ \hline
\rowcolor{lightgray} 
RP2: $\{15, 31, 128, 32 \mp j\}$      & 20.042              & 565                 & 29.852                  & 598          & 61,008,000 \\ \hline \hline
H: $\{512, 511, 513\}$                & 23.555              & 415+1\tnote{*}      & 31.863                  & 751          & 134,217,216 \\ \hline
RP1: $\{31, 32, 33, 64 \mp j\}$       & 21.09               & 775                 & 34.119                  & 720          & 134,119,392 \\ \hline
\rowcolor{lightgray} 
RP2: $\{15, 31, 128, 64 \mp j\}$      & 20.906              & 721                 & 33.27                   & 696          & 243,853,440 \\ \hline
\end{tabular}
\begin{tablenotes}
\footnotesize
\item[*] DSP.
\end{tablenotes}
\end{threeparttable}
}
\label{tab:7}
\end{table}

\begin{table}[H]
\centering
\caption{Moduli-Set Replacements for RES-DNN \cite{R17}}
\resizebox{\columnwidth}{!}{
\begin{tabular}{|c|c|c|c|c|c|}
\hline
\multirow{2}{*}{\textbf{Moduli-Sets}} & \multicolumn{5}{c|}{\textbf{Adder}} \\ \cline{2-6}
& \textbf{Delay (ns)} & \textbf{Area (LUT)} & \textbf{Power ($\mu$W)} & \textbf{PDP} & \textbf{DR} \\ \hline
$\{31, 128, 511\}$                    & 12.562              & 60                  & 17.356                  & 217          & 2,027,648   \\ \hline
RP1: $\{15, 17, 32, 16 \mp j\}$       & 12.257              & 65                  & 21.347                  & 262          & 2,097,120   \\ \hline
\rowcolor{lightgray} 
RP2: $\{15, 31, 32, 16 \mp j\}$       & 8.99                & 55                  & 20.191                  & 182          & 3,824,160   \\ \hline \hline
& \multicolumn{5}{c|}{\textbf{Multiplier}} \\ \hline
$\{31, 128, 511\}$                    & 23.126              & 279                 & 21.822                  & 505          & 2,027,648   \\ \hline
RP1: $\{15, 17, 32, 16 \mp j\}$       & 18.102              & 441                 & 24.505                  & 444          & 2,097,120   \\ \hline
\rowcolor{lightgray} 
RP2: $\{15, 31, 32, 16 \mp j\}$       & 18.559              & 430                 & 21.633                  & 401          & 3,824,160   \\ \hline
\end{tabular}
}
\label{tab:8}
\end{table}

Tables \ref{tab:3} to \ref{tab:8} illustrate that in certain instances, the area consumption of the proposed replacement moduli (RP1 and RP2) exceeds that of the reference counterparts. 
This phenomenon may be attributed to the magnitude of $n$, as evidenced by the fact that increasing $n$ leads to a more pronounced increase in the area of the proposed multiplier. 
With larger values of $n$, the synthesis tool faces challenges in optimizing the design, resulting in the dispersion of LUTs across the FPGA and consequently higher area utilization in terms of the LUT count. 
However, despite the increased area, the proposed moduli set consistently outperforms its predecessors in terms of delay and Power-Delay Product (PDP).


\begin{thebibliography}{10}

\bibitem{R1}
Rei Ueno and Naofumi Homma.
\newblock High-speed hardware architecture for post-quantum diffie–hellman key exchange based on residue number system.
\newblock In {\em 2022 IEEE International Symposium on Circuits and Systems (ISCAS)}, pages 2107--2111, 2022.

\bibitem{R2}
Ghassem Jaberipur and Bardia Nadimi.
\newblock Balanced $(3+2\log n)\delta g$ adders for moduli set $\{{2}^{n+1},2^{n}+2^{n-1}-1,2^{n+1}-1\}$.
\newblock {\em IEEE Transactions on Circuits and Systems I: Regular Papers}, 67(4):1368--1377, 2020.

\bibitem{R3}
Nikolai~I. Chervyakov, Pavel~A. Lyakhov, Nikolai~N. Nagornov, Dmitrii~I. Kaplun, Alexander~S. Voznesenskiy, and Danil~V. Bogayevskiy.
\newblock Implementation of smoothing image filtering in the residue number system.
\newblock In {\em 2019 8th Mediterranean Conference on Embedded Computing (MECO)}, pages 1--4, 2019.

\bibitem{R4}
Sahand Salamat, Mohsen Imani, Sarangh Gupta, and Tajana Rosing.
\newblock Rnsnet: In-memory neural network acceleration using residue number system.
\newblock In {\em 2018 IEEE International Conference on Rebooting Computing (ICRC)}, pages 1--12, 2018.

\bibitem{R5}
Nasim Samimi, Mehdi Kamal, Ali Afzali-Kusha, and Massoud Pedram.
\newblock Res-dnn: A residue number system-based dnn accelerator unit.
\newblock {\em IEEE Transactions on Circuits and Systems I: Regular Papers}, 67(2):658--671, 2020.

\bibitem{R6}
Zhi-Gang Liu and Matthew Mattina.
\newblock Efficient residue number system based winograd convolution.
\newblock In Andrea Vedaldi, Horst Bischof, Thomas Brox, and Jan-Michael Frahm, editors, {\em Computer Vision -- ECCV 2020}, pages 53--68, Cham, 2020. Springer International Publishing.

\bibitem{R7}
Vasilis Sakellariou, Vassilis Paliouras, Ioannis Kouretas, Hani Saleh, and Thanos Stouraitis.
\newblock On reducing the number of multiplications in rns-based cnn accelerators.
\newblock In {\em 2021 28th IEEE International Conference on Electronics, Circuits, and Systems (ICECS)}, pages 1--6, 2021.

\bibitem{R8}
Maria Valueva, Georgii Valuev, Mikhail Babenko, Andrei Tchernykh, and Jorge~M. Cortes-Mendoza.
\newblock Method for convolutional neural network hardware implementation based on a residue number system.
\newblock In {\em Programming and Computer Software}, volume~48, page 735–744, 2022.

\bibitem{R9}
Arman Roohi, MohammadReza Taheri, Shaahin Angizi, and Deliang Fan.
\newblock Rnsim: Efficient deep neural network accelerator using residue number systems.
\newblock In {\em 2021 IEEE/ACM International Conference On Computer Aided Design (ICCAD)}, pages 1--9, 2021.

\bibitem{R10}
Sahand Salamat, Sumiran Shubhi, Behnam Khaleghi, and Tajana Rosing.
\newblock Residue-net: Multiplication-free neural network by in-situ no-loss migration to residue number systems.
\newblock In {\em 2021 26th Asia and South Pacific Design Automation Conference (ASP-DAC)}, pages 222--228, 2021.

\bibitem{R11}
Valentina Arrigoni, Beatrice Rossi, Pasqualina Fragneto, and Giuseppe~S. Desoli.
\newblock Approximate operations in convolutional neural networks with rns data representation.
\newblock In {\em The European Symposium on Artificial Neural Networks}, 2017.

\bibitem{R12}
Hiroki Nakahara and Tsutomu Sasao.
\newblock A high-speed low-power deep neural network on an fpga based on the nested rns: Applied to an object detector.
\newblock In {\em 2018 IEEE International Symposium on Circuits and Systems (ISCAS)}, pages 1--5, 2018.

\bibitem{R13}
Wan-Ju Huang, Hsiao-Wen Fu, and Tsung-Chu Huang.
\newblock An-hrns: An-coded hierarchical residue number system for reliable neural network accelerators.
\newblock In {\em 2022 IEEE 31st Asian Test Symposium (ATS)}, pages 132--137, 2022.

\bibitem{R14}
Xilinx.
\newblock Sp701 evaluation kit board user guide (document no. ug1319).
\newblock Retrieved from [\url{https://www.xilinx.com/support/documents/boards_and_kits/sp701/ug1319-sp701-eval-bd.pdf}], 2019.

\bibitem{R15}
Yuke Wang.
\newblock Residue-to-binary converters based on new chinese remainder theorems.
\newblock {\em IEEE Transactions on Circuits and Systems II: Analog and Digital Signal Processing}, 47(3):197--205, 2000.

\bibitem{R16}
Shaahin Angizi, Arman Roohi, MohammadReza Taheri, and Deliang Fan.
\newblock Processing-in-memory acceleration of mac-based applications using residue number system: A comparative study.
\newblock In {\em Proceedings of the 2021 on Great Lakes Symposium on VLSI}, GLSVLSI '21, page 265–270, New York, NY, USA, 2021. Association for Computing Machinery.

\bibitem{R17}
Nikolai~I. Chervyakov, Pavel~A. Lyakhov, Maxim~A. Deryabin, Nikolay~N. Nagornov, Maria~V. Valueva, and Georgii~V. Valuev.
\newblock Residue number system-based solution for reducing the hardware cost of a convolutional neural network.
\newblock {\em Neurocomputing}, 407:439--453, 2020.

\bibitem{R18}
Chingyu Hung and Behrooz. Parhami.
\newblock An approximate sign detection method for residue numbers and its application to rns division.
\newblock {\em Computers \& Mathematics with Applications}, 27(4):23--35, 1994.

\end{thebibliography}

\noindent \textbf{Appendix}: \\
\label{appendix}
($F$-to-modulo-$(2^{2n}+1)$ conversion): Let $Z=2^{4n} Z_2+2^{2n} Z_1+Z_0$, represents the $5n-$bit numbers within the dynamic range of $F$, where $Z_2=z_{5n-1}\cdots z_{4n}\in [0,2^n-1]$, $Z_1=z_{4n-1}\cdots z_{2n}\in [0,2^{2n}-1]$, and $Z_0=z_{2n-1}\cdots z_0\in [0,2^{2n}-1]$. 
We obtain $X=|Z|_{2^{2n}+1}$, as follows. \\
$X=|Z|_{2^{2n}+1}=$ \\ 
$|2^{4n} Z_2+2^{2n} Z_1+Z_0 |_{2^{2n}+1}=|Z_2-Z_1+Z_0|_{2^{2n}+1}=$ 
$|Z_2+\overline{Z_1}-2^{2n}+1+Z_0|_{2^{2n}+1}=|Z_2+\overline{Z_1}+Z_0+2|_{2^{2n}+1}$. 
Let $U=u_{2n-1}\cdots u_0$, and $V=v_{2n-1}\cdots v_1 \overline{v_{2n}}$, be obtained, via a modulo-$(2^{2n}+1)$ carry-save adder operating on $|Z_2+\overline{Z_1}+Z_0+1|_{2^{2n}+1}$. 
Therefore, $X=|Z_2+\overline{Z_1}+Z_0+2|_{2^{2n}+1}=|U+V+1|_{2^{2n}+1}=|x_{2n} x_{2n-1}\cdots x_0+1|_{2^{2q}+1}$, where $|U+V|_{2^{2n}+1}$ is assumed to yield $x_{2n} x_{2n-1}\cdots x_0\in [0,2^{2n}]$, via a modulo-$(2^{2n}+1)$ adder. 
Finally, $X=|x_{2n} x_{2n-1}\cdots x_0+1|_{2^{2q}+1}=|x_{2n-1}\cdots x_0+1-x_{2n}|_2{^{2q}+1}=x_{2n-1}\cdots x_0+\overline{x_{2n}}\in [0,2^{2n}]$, where $x_{2n}=1\rightarrow x_{2n-1}\cdots x_0=0$.

\end{document}